\documentclass[aps,prl,twocolumn,groupedaddress]{revtex4-1}
\usepackage{graphicx}
\usepackage{dcolumn}
\usepackage{bm}
\usepackage{epsfig}

\begin{document}


\title{Intrinsic spin fluctuations reveal the dynamical response function of holes coupled to nuclear spin baths in (In,Ga)As quantum dots}

\author{Yan Li$^1$, N. Sinitsyn$^1$, D. L. Smith$^1$, D. Reuter$^2$, A. D. Wieck$^2$, D. R. Yakovlev$^{3,4}$, M. Bayer$^3$, S. A. Crooker$^{1*}$}

\affiliation{$^1$Los Alamos National Laboratory, Los Alamos, NM 87545, USA}
\affiliation{$^2$Angewandte Festk\"{o}rperphysik, Ruhr-Universit\"{a}t Bochum, D-44780 Bochum, Germany}
\affiliation{$^3$Experimentelle Physik 2, Technische Universit\"{a}t Dortmund, D-44221 Dortmund, Germany}
\affiliation{$^4$Ioffe Physical-Technical Institute, Russian Academy of Sciences, St. Petersburg 194021, Russia}

\date{\today}
\begin{abstract}
The problem of how single ``central" spins interact with a nuclear spin bath is essential for understanding decoherence and relaxation in many quantum systems, yet is highly nontrivial owing to the many-body couplings involved.  Different models yield widely varying timescales and dynamical responses (exponential, power-law, Gaussian, etc). Here we detect the small random fluctuations of central spins in thermal equilibrium (holes in singly-charged (In,Ga)As quantum dots) to reveal the timescales and functional form of bath-induced spin relaxation. This spin noise indicates long (400 ns) spin correlation times at zero magnetic field, that increase to $\sim$5 $\mu$s as hole-nuclear coupling is suppressed with small (100 G) applied fields. Concomitantly, the noise lineshape evolves from Lorentzian to power-law, indicating a crossover from exponential to inverse-log dynamics.
\end{abstract}
\maketitle

Single electron or hole ``central" spins confined in III-V semiconductor quantum dots (QDs) are promising candidates for solid-state qubits \cite{HansonRMP}. Although confinement suppresses momentum-dependent spin relaxation pathways, it enhances the hyperfine coupling between the central spin and the dense spin bath of $\sim$10$^5$ lattice nuclei comprising the QD. These hyperfine interactions dominate decoherence and spin relaxation at low temperatures. Within a QD ensemble, each central spin feels a different effective (Overhauser) magnetic field from nuclei, $\mathbf{B_n}$. Trivially, this leads to rapid nanosecond-timescale ensemble dephasing of an initially-oriented ensemble of central spins \cite{Merkulov}. On longer timescales, however, decoherence and relaxation of the central spin within \emph{each} QD occurs because $\mathbf{B_n}$ evolves slowly in time \cite{HansonRMP, Merkulov, Khaetskii, Erlingsson, Coish2004, Hassanieh, Chen, Yao, FischerPRB, Cywinski}. In large applied magnetic fields $B$, the huge difference between electronic and nuclear Zeeman energies suppresses flip-flop interactions between the two species, and $\mathbf{B_n}$ evolves primarily via weak dipolar coupling between nuclei. As $B$$\rightarrow$0, however, central spins can facilitate mutual ``co-flips" between distant nuclei (a process involving virtual flips of the central spin), which changes $\mathbf{B_n}$ more rapidly and accelerates spin relaxation \cite{HansonRMP, Merkulov, Khaetskii, Erlingsson, Coish2004, Hassanieh, Chen, Yao, FischerPRB, Cywinski}.

It is precisely in this low-field, intimately-coupled regime where the decoherence of the central spins becomes exceedingly difficult to model theoretically. Distinctly different timescales and a wide range of dynamical response functions (exponential, Gaussian, power-law) have been postulated, with exact solutions derived only under certain limiting assumptions, such as polarized nuclei. Low-field numerical models with unpolarized nuclei suggest interesting non-exponential dynamics with slow 1/log($t$) decays \cite{Coish2004, Erlingsson, Hassanieh, Chen}, highlighting the non-Markovian and strongly-correlated evolution of this many-body quantum system.

While groundbreaking experimental QD studies focused on electron central spins, considerable attention has recently shifted to holes \cite{Heiss, Gerardot, Brunner, EblePRL, Degreve, Kugler, Fras, FischerPRL, Trif}, whose \emph{p}-type wavefunctions avoid strong Fermi-contact hyperfine coupling to the lattice nuclei. Instead, hole-nuclear coupling occurs primarily via weaker dipolar (anisotropic hyperfine) interactions, reducing $\mathbf{B_n}$ by one order of magnitude \cite{Fallahi, Chekhovich, Degreve}. Optical studies of QD holes based on repeated initialization \cite{Heiss, Degreve, EblePRL, Kugler} or continuous pumping \cite{Gerardot, Brunner} have revealed long spin relaxation and coherence times in large $B$.  However, studies in the $B$$\rightarrow$0 limit, where hyperfine interactions are manifest most strongly, have received comparatively little attention \cite{Fras}. Moreover, the underlying functional form of the dynamical response of holes in a spin bath has not been explored.

As an alternative to conventional pump-probe techniques, the fluctuation-dissipation theorem suggests another route to reveal the dynamical response function of holes, that is based on passively detecting the spectrum of intrinsic random spin fluctuations of holes in thermal equilibrium (\emph{i.e.}, without optical pumping or initialization). This ``spin noise spectroscopy" has origins in atomic physics \cite{Aleksandrov, CrookerNature} and nuclear magnetic resonance \cite{Sleator}, and was subsequently demonstrated for electrons in bulk semiconductors \cite{OestreichPRL, CrookerPRB}. As recently applied to QDs \cite{CrookerPRL}, spin noise revealed the precession and rapid ensemble dephasing of holes in large $B$, due to an inhomogeneous distribution of hole g-factors. However, these noise-based methods have never been used to detect the intrinsic dynamics of central spins interacting with a nuclear spin bath.

\begin{figure}[t]
\includegraphics[width=.48\textwidth]{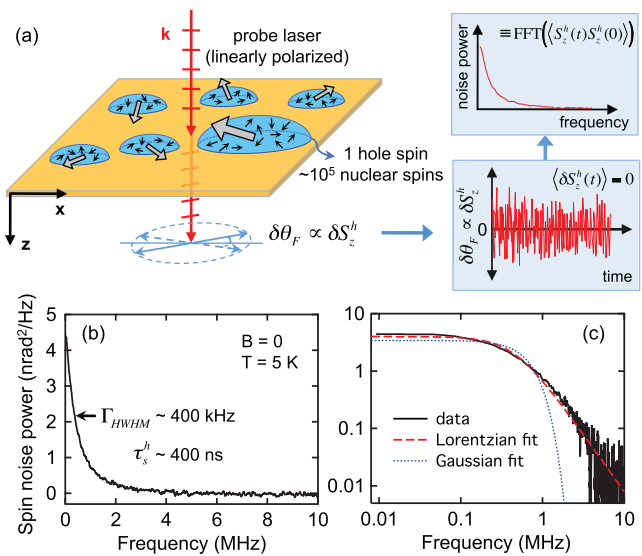}
\caption{(a) Experimental schematic: The random spin fluctuations $\delta S_z^h(t)$ of resident holes in (In,Ga)As/GaAs QDs impart Faraday rotation fluctuations $\delta\theta_F(t)$ on a linearly-polarized probe laser. The power spectral density of this ``spin noise" is measured with a balanced photodiode bridge and digital spectrum analyzer (see also \cite{Supp}). (b) Typical spin noise power spectrum of the resident holes at low temperature (5~K) and zero applied magnetic field ($B$=0). The 400~kHz half-width of the spin noise indicates a long $\sim$400~ns correlation time $\tau_s^h$ of the hole spins.  (c) The same spectrum on a log-log scale. The noise lineshape closely follows a Lorentzian, indicating exponentially-decaying hole spin correlations at $B$=0, in contrast with recent theories \cite{Coish2004, Erlingsson, Hassanieh, Chen}.} \label{fig1} \end{figure}

We therefore use a passive optical technique based on Faraday rotation to detect intrinsic hole spin fluctuations in (In,Ga)As QDs as $B$$\rightarrow$0, where coupling to the nuclear spin bath is most important. Crucially, because spin correlations are revealed in the spectral domain, this approach is well-suited to determine slow dynamical response functions with an accuracy sufficient to achieve a novel understanding of coupled spin-bath systems. In contrast with theoretical predictions \cite{Coish2004, Erlingsson, Hassanieh, Chen}, exponential dynamics with long (400~ns) correlation timescales are found at $B$=0. Using small (100~G) applied fields to suppress a dominant hole-nuclear interaction channel, even longer timescales of order 5 $\mu$s are revealed. Concomitantly, the fluctuation spectrum evolves from Lorentzian to power-law, indicating a crossover from exponential to inverse-log spin relaxation.

Figure 1(a) summarizes the spin noise experiment. A linearly-polarized probe laser is focused through an ensemble of weakly \emph{p}-type (In,Ga)As/GaAs QDs, where $\sim$10$\%$ of the QDs contain a single hole. Stochastic fluctuations of the ensemble hole spin projection along the sample normal \textbf{z}, $\delta S_z^h(t)$, impart Faraday rotation fluctuations $\delta \theta_F(t)$ on the probe laser via the usual optical selection rules for positively-charged trions. The power spectral density of $\delta \theta_F(t)$ is measured, revealing hole spin fluctuations in the frequency domain. Via the Wiener-Khinchin theorem, this is equivalent to the Fourier transform of the hole spin correlation function $\langle S_z^h(t)S_z^h(0)\rangle$. We emphasize that this passive technique probes the \emph{intrinsic} fluctuations of the resident hole spins \emph{while in thermal equilibrium} with the nuclear spin bath -- to leading order the detected spins are not optically pumped or excited \cite{Supp}, in marked contrast to conventional pump-probe methods. Moreover, since $\langle S_z^h(t) \rangle$ is always zero, only the intrinsic long-timescale dynamics of hole spins are detected in the zero-field limit. In contrast with prior work \cite{CrookerPRL}, two crucial differences are implemented here: i) low-noise stabilized probe lasers that permit accurate and quantitative recovery of the small low-frequency spin noise signals that exist in the zero-field limit, and ii) very low probe powers ($<0.5$~mW) and very large spot sizes ($\sim$100~$\mu$m) to ensure that we operate in a regime where the probe laser itself does not influence the measured spin noise signals and lineshapes (for details, see \cite{Supp}).

Using this approach, we find that the spectral density of this hole spin noise in zero applied field consists of a single, well-defined peak centered at zero frequency (Fig. 1(b)). Considerable information is encoded within this noise peak: Its half-width $\Gamma$ reveals the characteristic timescale $\tau_s^h$ of hole spin correlations $\langle S_z^h(t)S_z^h(0)\rangle$, and most importantly its detailed lineshape directly reveals \emph{the functional form} of the central (hole) spin decay -- a parameter of considerable theoretical interest \cite{HansonRMP, Merkulov, Khaetskii, Erlingsson, Coish2004, Hassanieh, Chen, Yao, FischerPRB, Cywinski}. On a log-log scale (Fig. 1(c)), we find this hole spin noise closely follows a Lorentzian lineshape over three orders of magnitude in frequency and signal, indicating that temporal correlations of $S_z^h(t)$ decay exponentially at $B$=0.  Note this contrasts directly with recent models predicting non-exponential dynamics in this regime \cite{Coish2004, Erlingsson, Hassanieh, Chen}.

Interestingly, the narrow 400~kHz width of the spin noise indicates very long hole spin correlation times of $\tau_s^h$$\sim$400~ns at $B$=0 ($\tau_s^h=1/(2\pi\Gamma)$ for exponential dynamics). $\tau_s^h$ greatly exceeds the $B$=0 hole relaxation time measured in InGaAs QDs \cite{EblePRL, Dahbashi} and is consistent with the hole relaxation recently inferred at $B$=0 from pump-probe studies of InAs QD ensembles \cite{Fras}. $\tau_s^h$ also exceeds the dephasing time of holes localized in narrow GaAs quantum wells studied by resonant spin amplification \cite{Kugler}.

\begin{figure}[tbp]
\includegraphics[width=.3\textwidth]{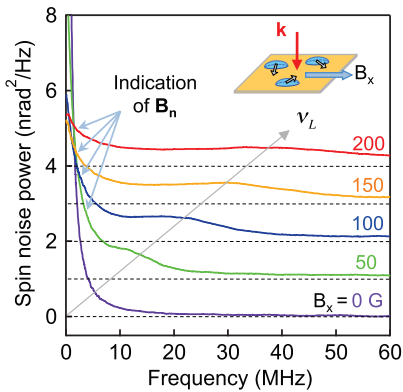}
\caption{Hole spin noise in purely transverse applied magnetic fields ($B_x$). In addition to the expected shift of the hole spin noise to the hole Larmor frequency ($\nu_L=g_\perp^h \mu_B B_x/h$, with $g_\perp^h$$\sim$0.15), there remains a finite noise component at zero frequency.  This reveals the presence of the longitudinal (\textbf{z}) components of the nuclear Overhauser magnetic field, $\mathbf{B_n}$.  Longitudinal fields, real or effective, necessarily result in spin noise at zero frequency \cite{Supp}. The integrated noise power remains constant.} \label{fig2} \end{figure}

The presence of hole-nuclear coupling becomes plainly evident upon applying small \emph{transverse} fields $B_x$ (see Fig. 2). As observed previously \cite{CrookerPRL}, the noise spectrum largely shifts to higher frequencies as fluctuations $\delta S_z^h$ are forced to precess about $B_x$ at the hole Larmor frequency. More importantly, however, Fig. 2 also reveals that a portion of the zero-frequency spin noise peak remains \emph{despite} application of purely transverse fields. This indicates that the holes do feel effective nuclear magnetic fields in the \textbf{z} direction, because longitudinal fields (real or effective) necessarily generate noise at zero frequency. In general, spins in an arbitrary magnetic field generate \emph{two} noise peaks: One at high frequency due to trivial spin precession, and one centered at zero frequency due to longitudinal field components. The former is weak at $B$=0 (for holes) and is strongly broadened due ensemble averaging, while the latter is not (for details, see \cite{Supp}). It is precisely this zero-frequency noise peak that we study, as it reveals the intrinsic timescales of $\langle S_z^h(t)S_z^h(0)\rangle$ and the dynamical response function of the hole spin decay.

\begin{figure}[tbp]
\includegraphics[width=.48\textwidth]{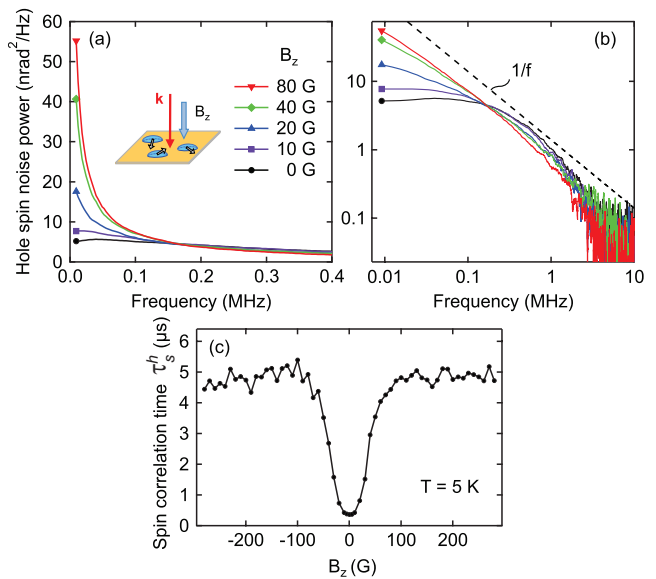}
\caption{(a) Hole spin noise spectra in purely longitudinal applied fields $B_z$=0, 10, 20, 40, 80 G. The spin noise narrows dramatically, indicating over an order-of-magnitude increase in hole spin correlation time $\tau_s^h$ from 400~ns to $\sim$5 $\mu$s. (b) The same spectra on a log-log scale clearly evolve away from a Lorentzian lineshape and more closely approach a 1/frequency power-law, indicating slow inverse-log spin decays in the time domain. The integrated noise power remains constant (as expected). (c) $\tau_s^h$ as a function of $|B_z|$.} \label{fig3} \end{figure}

To explore the extent to which hyperfine interactions limit $\tau_s^h$ at $B$=0, we apply small \emph{longitudinal} magnetic fields $B_z$ to overwhelm $\mathbf{B_n}$ and suppress hole-nuclear coupling. Figures 3(a,b) shows the spectral density of hole spin noise as $B_z$ increases to 80~G. The width of the noise peak narrows dramatically from 400~kHz to less than 40~kHz, indicating that $\tau_s^h$ increases over ten-fold to nearly 5 $\mu$s. More importantly, the detailed lineshape of the spin noise evolves away from Lorentzian and more closely approaches a $1/f$ power-law decay over the measured frequency range, thereby revealing an apparent crossover from exponential dynamics to much slower 1/log($t$) spin decays in the time domain. Higher fields to 300~G do not alter the noise lineshape further.

These data highlight an essential aspect of spin noise measurements: the ability to directly reveal detailed spectral lineshapes to explore slow and non-trivial decay mechanisms. The data appear to contradict recent theories \cite{Khaetskii, Erlingsson, Coish2004, Hassanieh, Chen} predicting slow, inverse-log decays of central \emph{electron} spins coupled to nuclear spin baths at $B$=0 (we see exponential decays at $B$=0). Whether these theories are fully applicable to holes remains an open question. We \emph{do} observe 1/$f$ noise spectra and inverse-log dynamics emerging in \emph{finite} (but small) $B_z$ however, suggesting at least partial validity of these models. Hole spin decays of order 1 $\mu$s can arise within models of hole-mediated nuclear co-flips, but robust Lorentzian noise lineshapes at $B$=0 are not expected (for details, see \cite{Supp}).  One possibility is that quadrupolar nuclear interactions and the local electric fields in (In,Ga)As QDs could rapidly `stir' fluctuations of $\mathbf{B_n}$ at $B$=0, accelerating hole relaxation via the co-flip mechanism and leading to exponential decays. Two-phonon spin relaxation processes \cite{Trif} or hybridization of hole states \cite{FischerPRL} have also been proposed, but their influence is not explicitly studied here. Note that these ensemble studies do not distinguish whether, in this regime, every hole in the ensemble exhibits 1/log($t$) dynamics, or whether there exists a broad distribution of, \emph{e.g.}, exponential timescales in the hole ensemble (due to varying QD strain and anisotropy), whose sum could mimic 1/log($t$) dynamics. Ongoing efforts are aimed at elucidating this difference.

\begin{figure}[tbp]
\includegraphics[width=.48\textwidth]{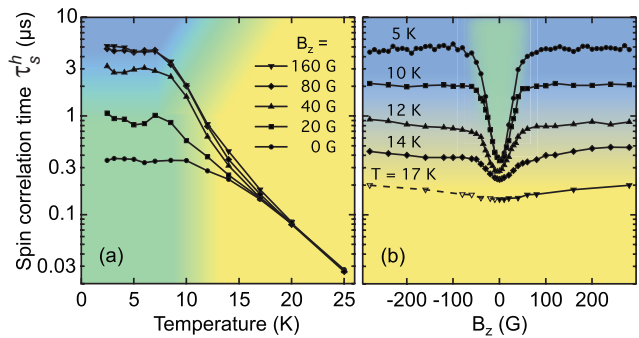}
\caption{Identifying different temperature and field regimes of hole spin decay. Data are plotted as (a) hole spin correlation time $\tau_s^h$ versus temperature at various applied fields $B_z$, and as (b) $\tau_s^h$ versus $B_z$ at various temperatures. Three general regimes can be identified (see text).} \label{fig4} \end{figure}

Although not strictly applicable to power-law dynamics, if we continue to infer a characteristic timescale $\tau_s^h$ via the noise half-width over the measured frequency range, then Figure 3(c) shows that $\tau_s^h$ increases rapidly with increasing $|B_z|$, but saturates at 5~$\mu$s when $|B_z|>100$~G. These data directly reveal the typical field scale of hole-nuclear coupling, $\sim$25~G, or about one-tenth of the electron-nuclear coupling in similar QDs \cite{Yakovlev}, consistent with previous studies \cite{Chekhovich, Fallahi, Degreve}.

Finally, Figure 4 shows a comprehensive study of how temperature and $B_z$ determine $\tau_s^h$ and the dynamical response function. The plots identify three general regimes: (i-yellow) At high temperatures $T\agt$ 15~K, $\tau_s^h$ falls rapidly, independent of $B_z$, and noise lineshapes are Lorentzian. This suggests straightforward phonon-assisted hole spin relaxation mediated by spin-orbit coupling, and exponential relaxation. (ii-green) Low temperatures below 10~K and in $B=0$, single phonon effects are suppressed, $\tau_s^h$ is limited to 400~ns by hole-nuclear coupling, and noise spectra remain unexpectedly Lorentzian. (iii-blue) Finally, at low temperature using small $|B_z|\agt$ 80~G to overwhelm $\mathbf{B_n}$, much longer 5 $\mu$s hole spin correlation times  are exposed and noise spectra approach $1/f$ power-laws, indicative of a crossover to very slow 1/log($t$) decays.

In summary we have demonstrated that spin noise spectroscopy allows unusually detailed studies, at all relevant timescales, of dynamic response functions in strongly-coupled hole-nuclear spin systems -- an inherently many-body problem that has eluded concise theoretical treatment. Systematic dependencies on temperature and magnetic field are revealed, serving as a test-bed for theoretical models. The measurement scheme of passively detecting the intrinsic spin fluctuations represents a kind of `quantum simulator', which is of great relevance to other interacting many-body systems of current interest including microcavity polariton condensates and fractional quantum Hall phenomena.

We thank M. M. Glazov, \L. Cywi\'{n}ski, I. \v{Z}uti\'{c}, J. Fabian, and F. Anders for helpful discussions. This work was supported by the Los Alamos LDRD program.

\onecolumngrid
\appendix

\newpage
\textbf{\Large{Supplemental Material}}

\renewcommand{\figurename}{FIG}
\renewcommand{\thefigure}{}

\section{A. Materials and Methods}

\textbf{Quantum dot structures.} InAs/GaAs QDs are grown by molecular-beam epitaxy on (001) GaAs substrates, and then thermally annealed at 940 $^\circ$C \cite{Yakov}. Annealing interdiffuses indium and gallium, resulting in (In,Ga)As/GaAs QDs with effective localization volumes of order 2000 nm$^3$, giving $\sim$10$^5$ nuclei within the spatial extent of the resident hole's wavefunction. The sample contains 20 layers of QDs, separated by 60~nm GaAs barriers. Each layer contains $\sim$$10^{10}$ QDs/cm$^2$. Although not intentionally doped, these QDs are weakly \emph{p}-type due to background carbon doping; we estimate that $\sim$10\% of the QDs contain a single resident hole. The inhomogeneously-broadened photoluminescence (PL) spectrum of these QD ensembles is typically peaked at $\sim$1.385~eV; see Fig. S1.

\textbf{Spin noise spectroscopy.} The QD samples are mounted on the cold finger of a small optical cryostat.  A linearly-polarized continuous-wave probe laser is tuned in wavelength to within the PL spectrum of the QD ensemble and is weakly focused through the sample ($\mathbf{k}\|\mathbf{z}\|\mathbf{n}$, where \textbf{n} is the sample normal). Stochastic fluctuations of the ensemble hole spin projection along the \textbf{z} axis, $\delta S_z^h(t)$, impart Faraday rotation fluctuations $\delta \theta_F(t)$ on the transmitted probe laser via the usual optical selection rules for positively-charged trions. Balanced photodiodes detect $\delta\theta_F(t)$, and the amplified output voltage $\delta V(t)$ is continuously digitized and Fourier-transformed to obtain the frequency spectrum of the measured noise power \cite{CrookPRL}. External coils provide longitudinal ($B_z$) and transverse ($B_x$) applied magnetic fields. Background noise densities due to photon shot noise and amplifier noise are eliminated by interleaving and subtracting spectra acquired at large $B_x$ ($>$2000 G), which shifts any spin noise to high frequencies outside the measured range. This procedure leaves behind only the noise signals arising from fluctuating hole spins at low fields.  Typically the cw probe laser power is a few hundred $\mu$W, and it is focused to a rather large (100 $\mu$m) spot on the sample to minimize heating and self-pumping of the QDs (see Fig. S2). Crucially, and in comparison with previous work \cite{CrookPRB,CrookPRL}, the present setup uses low-noise, stabilized probe lasers that now permit accurate and quantitative recovery of the small low-frequency spin noise signals that exist in the zero-field limit.

A consistent measure of the characteristic timescale $\tau_s^h$ of the hole spin correlations is obtained from the measured half-width $\Gamma$ of the spin noise peak that is centered on zero frequency.  Specifically, we use $\tau_s^h = 1/(2 \pi \Gamma)$, which is precise for Lorentzian noise lineshapes that indicate single-exponential relaxation dynamics.  This definition of $\tau_s^h$ is also used when the noise lineshapes deviate from Lorentzian, even though, strictly speaking, power-law lineshapes cannot be characterized by a specific timescale. In this case, $\Gamma$ is determined relative to the peak spin noise power spectral density that is measured in the lowest frequency bin.

\section{B. Dependence of integrated spin noise on photon energy of probe laser}

Figure S1 shows the inhomogeneously-broadened photoluminescence (PL) spectra of the (In,Ga)As/GaAs QD ensemble (solid black lines) under very low excitation conditions by a 1.58~eV (785~nm) laser. This PL arises from ground-state recombination of both positively-charged trions $X^+$ (from QDs containing a single resident hole), as well as from neutral excitons $X^0$ (from QDs that are empty). Note that these transition energies are typically very close in (In,Ga)As QDs ($X^+$ being higher in energy by $\sim2$ meV \cite{Greve}) and therefore they overlap in this ensemble PL spectrum and cannot be separately resolved. The PL spectrum therefore directly reflects the inhomogeneously-broadened distribution of fundamental QD transition energies in the ensemble.

\begin{figure}[h]
\includegraphics[width=.70\textwidth]{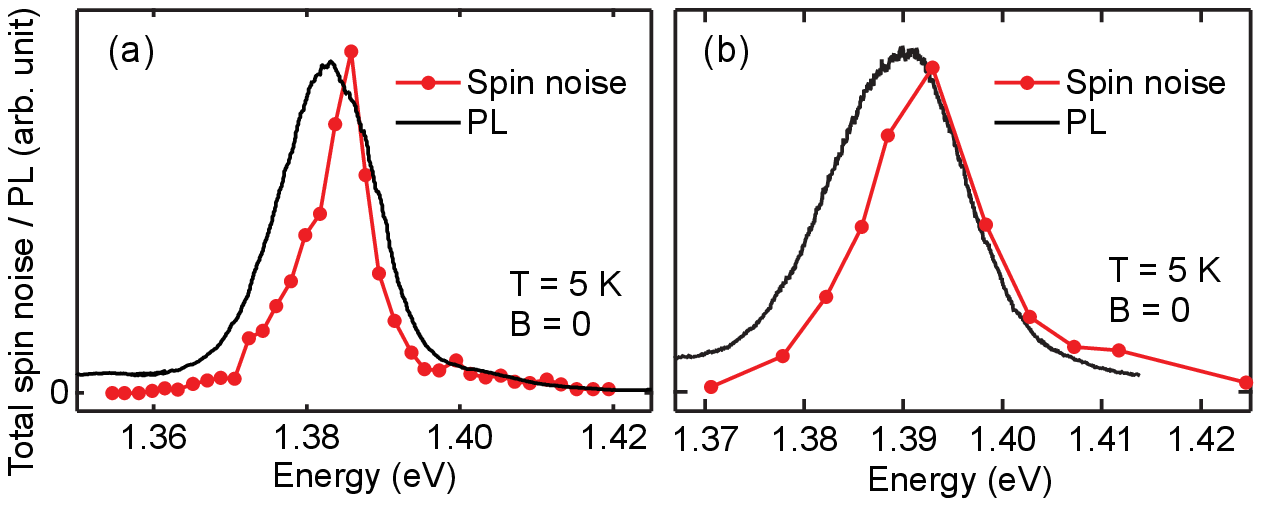}
\caption{S1. Dependence of the frequency-integrated (\emph{i.e.} total) spin noise power on the photon energy of the probe laser (points).  Also plotted is the photoluminescence (PL) spectra of the (In,Ga)As/GaAs QD ensemble (solid line). (a) and (b) show data from two different QD ensembles (both annealed at 940 $^\circ$C). The good correspondence between the PL and total spin noise indicates that the spin noise signals arise from resident holes in these QDs, and not from holes or electrons residing in, \emph{e.g.}, the wetting or buffer layers of the structure.} \label{S1}
 \end{figure}

When performing spin noise spectroscopy of these QDs, the narrow-band, continuous-wave probe laser is tuned in energy to within this PL spectrum. Figure S1 also shows the frequency-integrated (\emph{i.e.}, total) measured spin noise power as a function of the photon energy of the probe laser.  The integrated spin noise power provides a relative measure of the number of fluctuating spins being measured.  Its dependence largely follows the PL spectrum with a small blueshift, commensurate with the expected energy difference between $X^+$ and $X^0$ transition energies. This correspondence indicates that the measured spin noise arises from the resident holes that are trapped in the singly-charged subset of the QDs (rather than from spins in, \emph{e.g.}, the buffer or wetting layers or in the bulk of the semiconductor wafer). Further, at all probe laser energies where spin noise is detected, the spin noise exhibits the same narrow spectral width at zero applied magnetic field (as shown in Figure 1 of the main text), and the measured spin noise has the same behavior in transverse and longitudinal fields as shown in the main text (verified for a number of different probe energies).

\section{C. Spin noise spectroscopy: a passive measurement of hole spin fluctuations}

The particular implementation of spin noise spectroscopy employed in these experiments to detect fluctuations of $S_z^h$ (the net spin polarization of the resident holes in the QD ensemble) is based on optical Faraday rotation.  The Faraday rotation angle $\theta_F$ depends on the difference between the indices of refraction for right- and left-circularly polarized light, $n^+$ and $n^-$.  In particular, $\theta_F(\omega) = \frac{\omega L}{2c} [n^+(\omega) - n^-(\omega)]$, where $L$ is the sample thickness, $\omega$ denotes energy, and $c$ is the speed of light.

Before discussing the case of an inhomogeneously-broadened QD ensemble, first consider a spin noise measurement of a single homogeneously-broadened optical absorption resonance $\alpha(\omega)$ having a Lorentzian line-shape centered at $\omega_0$ and half-width $\gamma$; namely,  $\alpha(\omega) \sim \frac{\gamma}{(\omega - \omega_0)^2 + \gamma^2}$, as shown in Figure S2a. Let us say that this absorption resonance is spin-dependent as for the case of the optical transition between a resident hole and a positively-charged trion $X^+$.  Following the usual optical selection rules, if the resident hole is ``spin-up" then it can only absorb $\sigma^-$ circularly-polarized light, but if it is in the ``spin-down" state then it can only absorb $\sigma^+$ circularly-polarized light.

\begin{figure}[b]
\includegraphics[width=.65\textwidth]{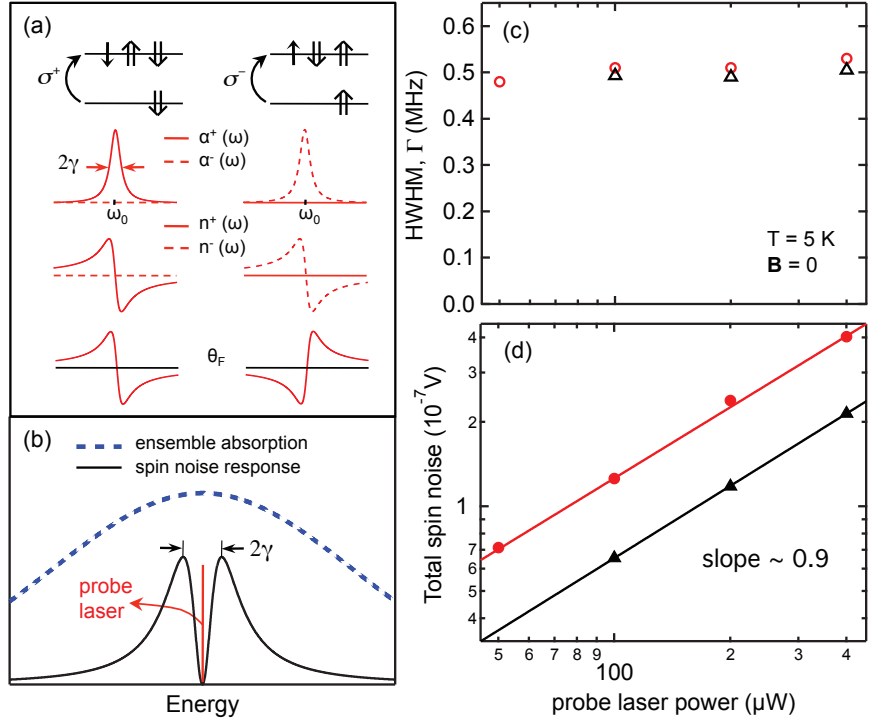}
\caption{S2. Spin noise spectroscopy using optical Faraday rotation: a non-perturbative measurement of resident hole spin fluctuations. (a) Illustrations depicting the $\sigma^+$ and $\sigma^-$ circularly polarized absorption resonances $\alpha^\pm (\omega)$ and the associated indices of refraction $n^\pm (\omega)$ of an idealized homogeneously-broadened (Lorentzian) optical transition (\emph{e.g.}, from a $X^+$ transition). $n^\pm (\omega)$ decay much more slowly than $\alpha^\pm (\omega)$ as a function of large detuning $\Delta$. Faraday rotation, $\theta_F \propto n^+ - n^-$, can therefore remain sensitive to the spin state $S_z^h$ of the resident hole, even for large $\Delta$  where $\alpha^\pm \rightarrow 0$. Importantly, note that $\theta_F=0$ when $\Delta =0$.  (b) An illustration of the inhomogeneously-broadened distribution of $X^+$ transitions within the QD ensemble. The probe laser is most sensitive to hole spin fluctuations in those QDs having $X^+$ transition energies detuned by about $\gamma$ and is \emph{not} sensitive to those QDs that are on resonance (\emph{i.e.}, to those QDs that are unavoidably pumped by the linearly-polarized probe laser). (c) The half-width $\Gamma$ of the zero-field hole spin noise spectrum measured using different probe laser power: There is no perceptible influence of the probe laser on the hole spin correlation time. (d) The total (frequency-integrated) spin noise, in volts, as a function of probe laser power. The total spin noise scales nearly linearly with the probe laser power, indicating no contribution to the spin noise from excited carriers. The dots are measured values, and the solid lines are power-law fittings. In (c) and (d) the red and black dots denote two measurements with different spot sizes.}\label{S2} \end{figure}

The dispersive part of this optical transition -- that is, the part complementary to the absorption resonances $\alpha^\pm(\omega)$ -- are the indices of refraction $n^\pm(\omega) \sim \frac{\omega - \omega_0}{(\omega - \omega_0)^2 + \gamma^2}$. As a function of detuning $\Delta = \omega - \omega_0$ from resonance, the indices $n^\pm(\omega)$ decay much more slowly than the absorption $\alpha^\pm(\omega)$: as $\Delta^{-1}$ versus $\Delta^{-2}$, respectively.  Therefore, a measurement of Faraday rotation can remain sensitive to the spin state of the hole even for large detuning $\Delta \gg \gamma$ where $\alpha^\pm(\omega) \rightarrow 0$ and the number of photons absorbed by the system becomes vanishingly small. In this regard, the measurement of $S_z^h$ can be considered non-perturbative, and spin noise measurements of this type were performed on alkali vapors at large probe laser detunings from the $S\rightarrow P$ (D1 and D2) lines of potassium and rubidium \cite{CrookNature}, and were also performed on conduction-band electrons in bulk \emph{n}-GaAs with the probe laser detuned well below the GaAs band-edge \cite{OestreiPRL, CrookPRB}.

Note that $\theta_F$ and therefore the spin noise measurement is most sensitive to spin fluctuations of this idealized system when the probe laser and the absorption resonance are separated by an energy of $\Delta \sim \gamma$.  The sensitivity falls off slowly as $\Delta^{-1}$ at larger detunings.  Even more importantly, note that $\theta_F \rightarrow 0$ at small detunings $\Delta <\gamma$, and that $\theta_F = 0$ exactly on resonance ($\Delta =0$) because there is no difference between $n^+(\omega)$ and $n^-(\omega)$.  \emph{The spin noise measurement is not sensitive to transitions lying at the same energy as the probe laser.}

This brings us to the case of measuring spin noise in an inhomogeneously-broadened QD ensemble. Here we tune the laser to directly within the inhomogeneously-broadened distribution of QD transition energies. As shown above in Figure S1, this maximizes the measured spin noise signal.  Undoubtedly, those QDs in the ensemble that happen to be resonant with the probe laser are optically pumped by the probe laser. However as noted in the previous paragraph, it is precisely these resonant QDs that do not, to leading order, give any spin noise (moreover, any photogenerated electrons, holes and/or trions that are pumped by the linearly-polarized probe laser are not spin-polarized). Rather, the probe laser will be primarily sensitive to those resident holes in QDs that have detuned $X^+$ transitions, and these QDs are not pumped (see Fig. S2b). In this regard, the Faraday rotation measurement still functions as a non-perturbative probe of $S_z^h$, the net spin polarization of resident holes in the QD ensemble.

The fact that $\theta_F (\omega=\omega_0)$ is not sensitive to spins in QD ensembles that are weakly pumped at $\omega=\omega_0$ was demonstrated in several recent ultrafast studies using independently-tunable pump and probe lasers \cite{Carter, Yugova, Glazov}.

Nonetheless, great care is taken to ensure that the probe laser functions only as a passive detector of the resident hole spin fluctuations, and does not inadvertently perturb the measurement of $S_z^h$.  Primarily, we wish to ensure that the probe laser i) does not heat the QDs which could lead to incorrect spin lifetime measurements, and ii) does not inadvertently detect any particles (electrons, holes, excitons or trions) that are photoexcited by the probe laser itself. Thus we use low probe laser intensities by using low laser power  ($\sim$200 $\mu$W) and large spot sizes ($\sim$100 $\mu$m). Although this significantly reduces the measured spin noise signals, it ensures that we operate in a regime where the measured spin correlation times $\tau_s^h$ (\emph{i.e.}, the inverse width of the spin noise spectra, $1/(2\pi \Gamma)$) are independent of probe laser power.

In this regime, Figure S2(c) shows that the measured half-width $\Gamma$ of the low temperature, zero-field hole spin noise spectrum is independent of the probe laser power. Significantly higher probe laser intensity leads to a broadening of the spin noise peak, indicating shorter hole spin lifetimes.  For the same series of measurements, Figure S2(d) shows the total (frequency-integrated) spin noise signal in units of volts of detected signal. As expected, the total spin noise signal detects no contribution from photoexcited carriers, since it increases nearly linearly with probe laser power (all else being equal, doubling the laser power simply doubles the voltages at the photodetectors). If the probe laser were inadvertently measuring spin noise from photogenerated (rather than resident) particles in the resonant QDs, the total spin noise would be expected to increase super-linearly (doubling the laser power would not only double the voltages at the photodetectors, it would also double -- or at least increase -- the number of particles being measured, leading to a superlinear dependence).

\section{D. Spin noise measurements of free electrons in n-type bulk Gallium Arsenide}

Figure S3 shows that magnetic fields in the longitudinal (\textbf{z}) direction, either real or effective, necessarily lead to some spin noise centered at zero frequency. To demonstrate this, we show spin noise measurements of electron-doped bulk GaAs. Free conduction band electrons in \emph{n}-type bulk GaAs are delocalized and sample a huge number of lattice nuclei. Therefore the influence of the fluctuating nuclear spin bath on these free electrons is extremely small.  To leading order, the only magnetic fields felt by the electrons are those that are externally applied. In this case, a purely transverse applied magnetic field $B_x$ uniformly shifts the spin noise of these electrons out to the Larmor frequency $\omega_L = g_e \mu_B B_x/\hbar$, and leaves no remnant of spin noise at zero frequency (Figure S3a).  This is in marked contrast to the spin noise spectra of QD holes shown in Figure 2 of the paper, in which some spin noise clearly remains at zero frequency despite application of a purely transverse $B_x$ -- this remaining spin noise is due to the longitudinal components of the effective nuclear field $\mathbf{B_n}$ that is felt by the holes.

\begin{figure}[h]
\includegraphics[width=.70\textwidth]{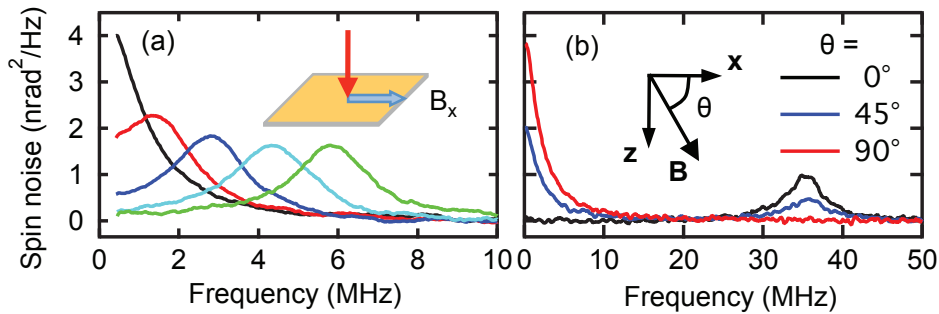}
\caption{S3. Spin noise from free electrons in bulk \emph{n}-type GaAs. (a) Electron spin noise in pure transverse fields $B_x$ = 0, 2.5, 5.0, 7.5, and 10~G (black to green). Note the absence of residual noise at 0 Hz, as expected from particles that feel little influence of the nuclear spin bath. The probe laser's wavelength is 845.6~nm (well below the GaAs band-edge), its power is 2~mW, and the spot size is large (100 $\mu$m). This $n$-GaAs wafer is 350 $\mu$m thick and is doped at $n_e = 1.4 \times 10^{16}$ cm$^{-3}$ (it is ``sample A" in Ref. \cite{CrookPRB}). (b) Spin noise in a 60~G canted applied magnetic field ($\mathbf{B} = |B|\textrm{cos} \theta~\mathbf{x}+ |B|\textrm{sin}\theta~ \mathbf{z}$), showing that longitudinal (\textbf{z}) field components lead to spin noise at zero frequency.} \label{S3} \end{figure}

That longitudinal fields generate spin noise at zero frequency is explicitly demonstrated in Figure S3(b), which shows electron spin noise spectra in an intentionally tilted applied magnetic field. A 60~G applied field \textbf{B} is rotated from the transverse to the longitudinal direction ($\theta=0\rightarrow \pi/2$). As \textbf{B} acquires a longitudinal component, zero frequency spin noise grows as sin$^2\theta$, with a commensurate cos$^2\theta$ suppression of noise signal at the electron Larmor frequency.

\section{E. Spin noise of holes in static nuclear fields}

It is useful to construct a toy model of hole spin noise in the considerably oversimplified limit of \emph{static} nuclear Overhauser fields $\mathbf{B_n} = (B_{n_x}, B_{n_y}, B_{n_z})$.  The purpose of this exercise is three-fold:

1) It demonstrates that hole spin precession about $\mathbf{B_n}$ generates a broad hole spin noise spectrum at high frequencies between 5-100~MHz.  The broadness of this high frequency noise is due to the statistical distribution of $\mathbf{B_n}$ over the QD ensemble, and also to the inhomogeneous distribution of hole g-factors within the QD ensemble.

2) It demonstrates that the longitudinal Overhauser fields $B_{n_z}$ give a delta-function (or at least very narrow) spin noise peak at zero frequency, that is \emph{not} expected to narrow or broaden with applied fields $B_x$ or $B_z$ (in contrast to actual experimental observation). This noise peak is not statistically broadened by the ensemble, since each QD gives some noise at exactly zero frequency.

3) It shows that, in zero applied field, the high-frequency precessional noise is strongly suppressed as compared to the zero-frequency noise due to the large anisotropy of the hole g-factor.

In marked contrast with electrons, holes couple very anisotropically to in-plane versus out-of-plane magnetic fields. For \emph{pure} heavy holes, the longitudinal (out-of-plane) g-factor $g_\parallel$ is finite while the transverse (in-plane) g-factor $g_\perp$ is zero.  However, hole eigenstates in typical \emph{p}-type self-assembled III-V quantum dots invariably contain some admixture of light hole states in addition to their predominantly heavy-hole character. This leads to a small in-plane g-factor $g_\perp$ that is of order 0.15 in our QDs, which is about an order of magnitude less than $g_\parallel$ ($g_\parallel \sim 1$). Note also that there exists a large inhomogeneous dispersion of these g-factors within the QD ensemble, likely due to differences in QD shape and strain.

Here we assume that the Overhauser field $\mathbf{B_n}$ in each quantum dot has components $B_{n_x}, B_{n_y}$, and $B_{n_z}$ that are each Gaussian-distributed with typical dispersion of $\sim$25 Gauss (taken from experimental data of Figure 3c). In any given QD, hole spin precession about the transverse component of $\mathbf{B_n}$ generates noise in $S_z^h$ (the measured quantity) at the hole's Larmor precession frequency,

\begin{equation}
\omega_L=\frac{\mu_B}{\hbar}\sqrt{g_\perp^2(B_x+B_{n_x})^2+g_\perp^2(B_{n_y})^2+g_\parallel^2(B_z+B_{n_z})^2}
\label{f}
\end{equation}
(where here we have generalized slightly to allow for real applied magnetic fields $B_x$ and/or $B_z$, and we assume no in-plane anisotropy of $g_\perp$ for simplicity).

In each QD, $\omega_L$ occurs at a different frequency depending on the magnitude and direction of $\mathbf{B_n}$ in that QD, and also depending on $g_\perp$ and $g_\parallel$ in that QD. Using the typical values stated above, $\omega_L$ can range from a few MHz out to $\sim$100~MHz.  Averaging over many QDs, \emph{this precessional noise generates a very broad spectrum}, weakly peaked at about 10~MHz and spanning the range from 5-100 MHz. It is this precessional noise that represents the trivial ensemble dephasing of an ensemble of hole spins that are all initially oriented at $t=0$, as in a pump-probe measurement.

\begin{figure}[h]
\includegraphics[width=.5\textwidth]{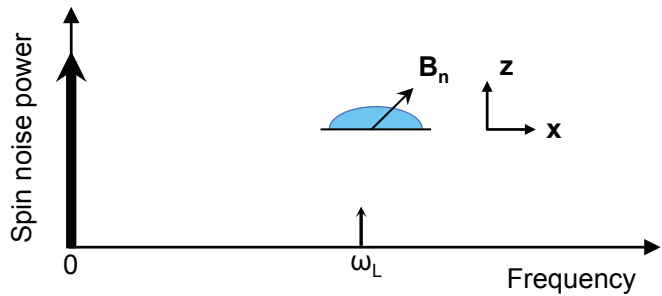}
\caption{S4. A toy model of hole spin noise in a QD, assuming a static nuclear field $\mathbf{B_n}$. In each QD, hole spin precession about $\mathbf{B_n}$ induces a high-frequency noise peak at the hole Larmor frequency $\omega_L$. In addition, longitudinal (\textbf{z}) components of $\mathbf{B_n}$ generate a noise peak at zero frequency (for the analogous case of electrons in arbitrary applied fields, see Fig. S3 above). For the purposes of this toy model, these peaks are represented by delta functions. The total noise power contained within the high-frequency peak is generally much smaller than that within the zero-frequency peak, due to the large anisotropy of the hole g-factor.  Statistical averaging over the QD ensemble smears the high-frequency precessional noise over a large frequency range.  On the other hand, the zero-frequency noise peak is contributed to by every QD.  \emph{In our noise experiments, it is precisely the width and lineshape of this zero-frequency peak that we seek to measure}, as these parameters reveal the intrinsic spin correlation timescales and decay mechanisms of the hole spins.} \label{S4} \end{figure}

However, in each and every QD, the longitudinal component of $\mathbf{B_n}$ (\emph{i.e.}, $B_{n_z}$), generates a finite time-averaged value of the hole's spin projection, $\langle S_z^h(t) \rangle$, that does \emph{not} decay in time. In a noise measurement, and within the limitations of this toy model, this simply gives a delta function at zero frequency, and applied magnetic fields do not alter the lineshape of this peak. Of course in reality, $\langle S_z^h(t) \rangle$ \emph{does} decay because $\mathbf{B_n}$ is \emph{not} static, and the associated noise peak is not a delta function -- \emph{it is precisely the width and lineshape of this zero-frequency peak that reveals the long-time decay mechanisms of the central hole spins} that we are interested in.

Within this simple model, it can be shown that the total (frequency-integrated) power of the noise peak at $\omega_L$ is
\begin{equation}
P_L=\frac{g_\perp^2(B_x+B_{n_x})^2+g_\perp^2(B_{n_y})^2}{g_\perp^2(B_x+B_{n_x})^2+g_\perp^2(B_{n_y})^2+g_\parallel^2(B_z+B_{n_z})^2}
\label{Af}
\end{equation}

while the integrated power of the noise peak at zero frequency is
\begin{equation}
P_0=\frac{g_\parallel^2(B_z+B_{n_z})^2}{g_\perp^2(B_x+B_{n_x})^2+g_\perp^2(B_{n_y})^2+g_\parallel^2(B_z+B_{n_z})^2}
\label{A0}.
\end{equation}

Therefore in the absence of any applied fields ($B_x = B_z =0$), the noise power at $\omega_L$ is typically suppressed as compared to the zero-frequency noise by $\sim (g_\perp / g_\parallel)^2$ (except when $B_{n_z}$ is very small). This, together with the above-noted fact that ensemble averaging smears out the high-frequency precessional noise over a very broad bandwidth, may explain why we do not observe any clear sign of precessional noise occurring at high frequencies in the absence of applied fields. Only the much larger zero-frequency noise is apparent. However, as we apply large magnetic fields in the transverse direction ($B_x$), all the hole noise power is expected to shift to the precessional noise component at $\omega_L$, in agreement with our experimental data (Figure 2 of the main paper). Further, as we apply large fields in the longitudinal direction ($B_z$), the total noise power contained within the zero-frequency component will increase only very slightly (all the noise power is essentially already contained in this peak), again in agreement with experimental observation.

Finally, note that the above expressions also hold for electrons, in which case $g_\perp \sim g_\parallel$ and the high-frequency and zero-frequency noise peaks are expected to have comparable integrated power.

\section{F. Contribution to the hole spin noise from hyperfine-mediated ``co-flips" of nuclear spins}

Previous theoretical studies of nuclear spin ``co-flips" (a flip-flop of two distant nuclei mediated by virtual hole-nuclear spin flips) revealed non-exponential decay of the central spin due to non-Markovian evolution and strongly-correlated dynamics \cite{glazman,dobrovitski}.

The goal of this section is to show that the time scale of $\sim$1~$\mu$s, which roughly corresponds to the inverse width of the spin noise power in zero magnetic field, is in agreement with a simple theory of how nuclear co-flips lead to central spin relaxation. We will also highlight differences between this theory and the observed behavior of noise power spectrum at longer time scales.

Consider the effective Hamiltonian of a single hole spin interacting with $N \sim 10^5$ nuclear spins, in the presence of an applied out-of-plane (longitudinal) magnetic field $B_z$.  We have
\begin{equation}
\hat{H}=  \hat{\omega}_z  \hat{S}_z+  \hat {V},
\label{Ham}
\end{equation}
\begin{equation}
\hat{V}\equiv \sum_i  \gamma_{\perp}^i \hat{s}_{ix} \hat{S}_x, \quad \hat{\omega}_z = g_\parallel B_z + \sum_i  \gamma_{||}^i \hat{s}_{iz},
\label{v}
\end{equation}
where $\hat{S}$ stands for the hole spin operator, $\hat{s}_{i}$ is the $i$-th nuclear spin operator, $\gamma_{||}$ and $\gamma_{\perp}$ are the coupling strengths, out-of-plane and in-plane respectively, between the central spin and nuclear spins. We assume, for simplicity, that all spins are $1/2$, and  $\gamma_{||}$ and $\gamma_{\perp}$ are in the same ratio as the ratio of longitudinal and transverse hole g-factors, $g_\parallel$ and $g_\perp$. Thus, $\gamma_{||}$ is about ten times greater than $\gamma_{\perp}$. Note that in general $\gamma^i_{\perp}$ is complex, reflecting the fact that transverse coupling involves both $x$ and $y$ spin components, but as this does not influence the following discussion we treat these couplings as real positive parameters.

The large value of $N$ and the coupling anisotropy enable a perturbative approach \cite{glazman}, in which the zeroth order wave function of a typical state can be approximated to be the eigenstate of the total spin operator along $z$-axis, for example,

\begin{equation}
\vert \Psi \rangle \approx |\Psi^0 \rangle = \vert \Uparrow, \downarrow \ldots \downarrow \uparrow \ldots  \uparrow \rangle,
\label{zero-order}
\end{equation}
where $\Uparrow$ or $\Downarrow$ indicates the state of the hole central spin (up or down along $z$-axis) and other arrows show states of the nuclear spins.
We will call the expectation value of operator $\hat{\omega}_z$ in the state $|\Psi \rangle$ the {\it bias}, and we will denote it by $\omega_z$. We will show that this part of the hyperfine coupling resists the flip-flop transitions.

Diffusive dynamics of the Overhauser field $\mathbf{B_n}$ are possible because the couplings of central spin to different nuclear spins are not equal so that exchanging direction of a pair of nuclear spins changes the bias value by the amount $\sim \gamma_{||}$. According to the Hamiltonian (\ref{Ham}), such pair-wise co-flips of nuclear spins happen due to transitions through the virtual states with a flipped central spin. The accumulation of pair-wise nuclear spin-flips leads to diffusion in the space of Overhauser field values. The expectation value of the central spin follows the direction of the Overhauser field.
When the bias changes across the region with  $B_z + B_{n_z} \alt (\gamma_\perp / \gamma_\parallel) \sqrt{\langle \mathbf{B_n}^2 \rangle}$, the variation of this direction becomes substantial, corresponding to the relaxation of the central spin expectation value.

There is no direct coupling between the state (\ref{zero-order}) and the state
\begin{equation}
\vert \Psi_{ij} \rangle = \vert \Uparrow,  \{ i, j\} \rangle,
\label{virt}
\end{equation}
where $\{i,j \}$ means that nuclear spins $i$ and $j$ change their directions in comparison to the state $\vert \Psi \rangle$. Therefore, to estimate the rate of nuclear spin co-flips we
 use first order perturbation theory to incorporate transitions to the virtual states in our wave functions:
\begin{equation}
\vert \Psi \rangle   \approx  C\left( |\Psi^0 \rangle - \sum_{i} \frac{  \gamma_{\perp}^i  } { 2\omega_z} \vert V_i^0 \rangle  \right)
\label{second}
\end{equation}
where
\begin{equation}
\vert V_i ^0\rangle = \vert \Downarrow,  \{ i \} \rangle,
\label{virt}
\end{equation}
and where $\{ i \}$ means that $i$-th nuclear spin is flipped from up to down state,
and $C$ ensures a normalization. This definition of a typical state $|\Psi \rangle$ includes fast transition processes to the virtual states  $\vert V_i ^0\rangle $. One can see that $\hat{V}$ already directly couples states $|\Psi \rangle$ and $|\Psi_{ij}\rangle$.

Further perturbation expansion breaks down because there are many states $|\Psi_{ij}\rangle$ with almost the same energy as the initial state $|\Psi\rangle$.
The presence of such  quasi-degenerate states leads to incoherent transitions from $\vert \Psi \rangle$ into one of the states $\vert \Psi_{ij} \rangle$. The rate $1/\tau$ of such transitions can be estimated by Fermi's Golden Rule as
follows: There are $N^2/4$ pairs of distinct states $ \vert \Psi_{ij} \rangle $ (again, we are considering spin 1/2 nuclei for convenience), distributed around the mean value with interval of energies $~ {\delta \gamma}_{||}$, where
${\delta \gamma}_{||}$ is the width of the distribution of $\gamma_{||}^i$ around their mean value.  Thus the rate, $\tau^{-1}$, of an incoherent transition from the initial state $|\Psi\rangle$ to one of the states $|\Psi_{ij}\rangle$ that differ from $|\Psi\rangle$ by  co-flips of two nuclear spins is given by

\begin{equation}
 \tau^{-1} \sim 2 \pi \left( \frac{\gamma_{\perp}^2}{ 4\omega_z }\right)^2 \frac{N^2}{4 \delta \gamma_{||}},
 \label{gam}
\end{equation}
where $N^2/(4 \delta \gamma_{||})$ is the density of states represented by all possible pairs $|\Psi_{ij} \rangle$, and $\gamma_{\perp}^2/{ 4\omega_z }$ is the strength of a typical coupling between states $|\Psi_{ij} \rangle$ and the state $|\Psi \rangle$, \emph{i.e.} $\langle\Psi_{ij}|\hat{V}|\Psi\rangle$.

We use  $\gamma_{||} \sim M \Omega_0 / \Omega$, where according to various estimates, $M \sim$ 3-13 $\mu$eV \cite{Testelin, Loss} is the hyperfine coupling per nuclear spin per unit probability of the hole being in the unit cell that contains this nuclear spin. The factor $\Omega_0 / \Omega$ is the ratio of volumes of the
unit cell $\Omega_0\sim$ (0.57~nm)$^3$ to the volume of the quantum dot, $\Omega \sim$ 2000~nm$^3$.  Hence $\Omega_0 / \Omega \sim 10^{-4}$ and
$\gamma_{||}\sim  1\cdot 10^{6}$ s$^{-1}$.

The fastest spin relaxation happens in states that have initial bias $\omega_z < \gamma_{\perp}\sqrt{N}/2$ because such states experience comparable couplings along longitudinal (along $z$-axis) and transverse axes.
In such states, changes in total field along $z$-axis (also the bias $\omega_z$) leads to a substantial change of the orientation of the total field that acts on the hole's spin. If we consider a state with initial $\omega_z$ close to $\omega_z \sim \gamma_{\perp}\sqrt{N}/2$  we find from (\ref{gam}) that the typical time of an incoherent transition between states that differ by a co-flip energy is  $ \tau  \sim  10^{-9}$ s.

This time $\tau$ should not be confused with a hole's spin relaxation time $\tau_s^h$ because a single co-flip does not lead to a substantial change of the total hyperfine field. Moreover, such random transitions can lead both to an increase or a decrease of $\omega_z$. However, accumulation of co-flip transitions leads to a diffusion in the space of bias values. Hence the latter changes with time according to the diffusion law:

$$
\sqrt{\langle (\delta \omega_z)^2 \rangle }  \sim \gamma_{||} \sqrt{t/\tau},
$$
where the coefficient $\gamma_{||}$ reflects the fact that a single co-flip corresponds to a change of the bias of order $\gamma_{||}$.
The hole's spin rotation angle becomes of order unity when the change of the bias becomes comparable to the initial bias, $\gamma_{\perp} \sqrt{N}/2$.
This gives us an estimate of the hole's spin relaxation time:
$$
\tau_s^h \sim \tau N (\gamma_{\perp}/\gamma_{||})^2 \sim 10^{-6}{\rm s} .
$$
The experimentally obtained value, $\tau_s^h \sim 400$ ns, is in very reasonable agreement with this estimate. Here we note, however, that the region of bias values with  $\omega_z < \gamma_{\perp}\sqrt{N}/2$, accounts for only about 10\% of statistically possible states of the nuclear spin bath. When $\omega_z$ exceeds the size of this region, incoherent flip-flop transition rate is suppressed, as it can be seen from the fact that $\omega_z$ enters as a second power in the denominator
in Equation (\ref{gam}). Thus, for  values of  $\sqrt{\langle \omega_z^2 \rangle} \sim \gamma_{||}\sqrt{N}/2$, the flip-flop transition rate is suppressed by a factor 100, in disagreement with the assumption of a single relaxation time for all quantum dots, which is expected from the Lorentzian shape of the power spectrum. Dipole interactions between nuclear and central spins allow additional coupling terms in the interaction Hamiltonian, such as $\sim \hat{S}_z \hat{s}_{ix}$  that could contribute to the nuclear spin dynamics, but such terms were estimated to be negligibly small for heavy-hole states in quantum dots \cite{Testelin}. It is also possible that strains and nonuniform doping introduce high gradients of local electric fields that couple to quadrupole moment  of nuclear spins. Quadrupole nuclear interactions with nonuniform electric field often broaden the solid state NMR-lineshapes of nuclei to the MHz range \cite{quad}. Such interactions do not directly involve the central spin but, if sufficiently strong, they can induce fast intrinsic dynamics of nuclear spins and stir fluctuations of the Overhauser field across its typical values. It is possible that, supplemented by fast incoherent co-flip processes near the zero bias values, such fluctuations lead to the relaxation of the central spin at a fraction of a microsecond.

Finally, we discuss consequences for the case of nonzero applied fields $B_z$.
According to our theory, the window of  Overhauser field values that allow fast spin relaxation due to co-flips corresponds to $|B_z+B_{n_z}|<\gamma_{\perp} \sqrt{N}$.
The average number of quantum dots that happen to have the hyperfine field inside this narrow window at  same
moment of time follows the Gaussian distribution of the hyperfine field $|\mathbf{ B_n}|$.  Consequently, this number is suppressed when $|B_z|>\sqrt{\langle \mathbf{B_n}^2 \rangle}$, while the relative contribution to power spectrum from the states having the bias far from the fast relaxation window increases with $B_z$.
Hence, the width of the noise power spectrum should quickly decrease with $B_z$ when $|B_z|>\sqrt{\langle \mathbf{B_n}^2 \rangle}$, in agreement with the experimental data.


\end{document}